# NUMERICAL STUDY OF THE CRYOGENIC COOLING OF AMPLIFIERS FOR HIGH POWER LASERS

Bellec M.[1]*, Luchier N.[1], Balarac G.[2], Bieder U.[3] and Girard A.[1]
*Author for correspondence
1. Univ. Grenoble Alpes, CEA, IRIG, DSBT, Grenoble, France
2. Laboratoire des écoulements géophysiques et industriels (LEGI) - CNRS-UGA-G-INP, Grenoble, France
3. CEA, Université Paris-Saclay, DEN-STMF, Gif-sur-Yvette, France
E-mail: morgane.bellec@cea.fr

**ABSTRACT**
The French collaborative Trio4CLF project aims to understand and control the cryogenic cooling of amplifiers for high power (~1 PetaWatt) and high repetition rate (1-10 Hertz) lasers. In such amplifiers, the fluid (low temperature gaseous helium) evacuates the thermal power absorbed by the solid amplifying plates. A precise knowledge of the heat exchange and of the turbulent fluid flow in the amplifier is requested to evaluate its impact on the laser beam quality. Large Eddy Simulations representative of the amplifier cooling are performed using TrioCFD, a code developed by the CEA. First, validation simulations are carried out for heated periodic channel flows, allowing comparisons with Direct Numerical Simulation results from the literature. They are conducted using conjugate heat transfer calculation between the fluid and the solid. The channel flow is turbulent, with a bulk Reynolds number (based and the bulk velocity and the total height of the channel) of about 14000. Large Eddy Simulation of a heated open turbulent cryogenic helium flow with developing thermal boundary layers is then presented.

**INTRODUCTION**

The thermal management of solid-state laser amplifiers has become a critical issue in order to increase the repetition rate of high power lasers [1]. A cryogenic cooling of the amplifier is especially interesting due to the better thermal and optical properties at low temperatures than at room temperature of the crystals used as amplification medium [2, 3]. To increase the heat transfer surface, the studied amplifier [4] has a slab configuration as shown in Fig. 1. It is cooled by a forced flow of gaseous helium.

In such configuration, the laser beam is perpendicular to the helium flow. Due to the low viscosity of cold helium, turbulence will develop in the flow. An inhomogeneous temperature field will produce inhomogeneity in the refractive index distribution through density variations. The effect on the laser beam coherence must be assessed. A similar problem occurs when a laser beam propagates through a compressible turbulent boundary layer, with the difference that the density variations are then induced by the pressure fluctuations rather than the temperature ones [5, 6]. Tromeur *et al.* [5] recommended the use of Large Eddy Simulations (LES) to correctly evaluate these aero-optical effects due to inaccuracies obtained when using optical models combined with RANS simulations. Mani *et al.* [6] confirmed that a typical LES will correctly capture those effects.

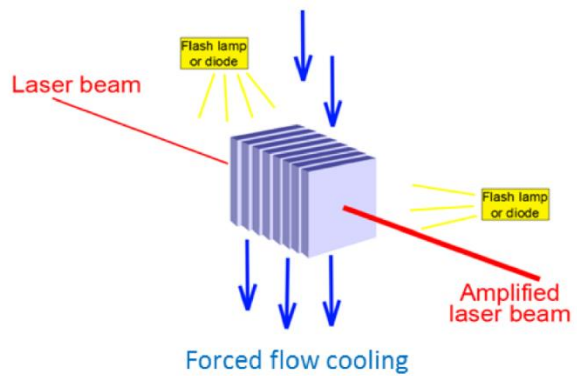

**Figure 1** Schematic view of the laser amplifier

Several numerical studies exist in the literature on fully developed turbulent channel flow with heat transfer. Kawamura *et al.* [7] conducted Direct Numerical Simulations (DNS) at different Prandtl and Reynolds numbers using a non-fluctuating temperature boundary condition at the walls, constituting a reference database. Bricteux *et al.* [8] conducted similar simulations in DNS and LES, in the specific case of a very low Prandtl number, typical for liquid metals. These studies overlooked the actual wall/fluid heat transfer by using ideal non-fluctuating temperature boundary condition. Tiselj *et al.* [9] conducted DNS of turbulent channel flow with conjugate heat transfer at different Reynolds numbers and for different solid and fluid properties. They showed that the thermal boundary condition has a small influence on the mean temperature and on the temperature heat fluxes, but that it affects the temperature fluctuations in the fluid. Conjugated heat-transfer is also used in more complex geometries such as channel flow with a wire embedded in one of the wall [10].

A step-by-step strategy is adopted. In the present study, LES are performed in conditions representative of a simplified cryogenically cooled laser amplifier, namely a turbulent channel flow of gaseous helium at 80 K between two crystal plates at 105 K. Validation simulations in a periodic channel are first presented to assess the quality of the simulations by comparison with DNS results of the literature. Since temperature fluctuations are the quantity of interest, the crucial prediction of the heat transfer between the solid and the fluid is calculated by conjugated heat transfer. These calculations revealed that Dirichlet boundary conditions with imposed temperature lead to similar results. A LES of a heated open channel flow is then conducted using such imposed temperature boundary conditions. The developing thermal boundary layers are presented.

### NOMENCLATURE

| Symbol | Units | Description |
|---|---|---|
| $C_p$ | [J/kgK] | Heat capacity |
| G | [m$^2$/kg] | Gladstone-Dale constant |
| h | [m] | Half-height of the channel |
| $h_{exch}$ | [W/m$^2$K] | Heat transfer coefficient |
| k | [rad/m] | Wave number |
| L | [m] | Domain dimension |
| m | [kg] | Mass |
| n | [-] | Index of refraction |
| N | [-] | Number of cells |
| Nu | [-] | Nusselt number |
| P | [Pa] | Pressure |
| Pr | [-] | Prandtl number |
| Q | [W] | Power |
| q | [K/s] | Source term |
| Re | [-] | Reynolds number |
| S | [m2] | Exchange surface |
| T | [K] | Temperature |
| t | [s] | Time |
| U | [m/s] | Velocity |
| V | [m$^3$] | Volume |
| x | [m] | Streamwise axis |
| y | [m] | Crosswise axis |
| z | [m] | Spanwise axis |

Special characters

| Symbol | Units | Description |
|---|---|---|
| α | [m$^2$/s] | Thermal diffusivity |
| λ | [W/mK] | Thermal conductivity |
| $λ_l$ | [m] | Wavelength |
| ρ | [kg/m$^3$] | Density |
| $\mathcal{T}$ | [s] | Time constant |
| ν | [m$^2$/s] | Kinematic viscosity |
| φ | [rad] | Wave phase |

Subscripts and superscripts

| | |
|---|---|
| b | Bulk |
| i | Any Cartesian direction |
| f | Fluid |
| RMS | Root Mean Square |
| s | Solid |
| τ | Friction |
| w | Wall |
| + | Normalization |

### NUMERICAL MODEL

All simulations were performed using TrioCFD [11], an open-source CFD code developed by the Nuclear Energy Division of CEA. Physical properties of the media, geometry of the calculation domain as well as thermohydraulic conditions of the calculations are representative of a cryogenically cooled laser amplifier.

At a temperature of T=80 K and a pressure of P=5×10$^5$ Pa, helium is a gas with a kinematic viscosity ν=3×10$^{-6}$ m$^2$/s. The half-height of the cooling channel, formed by the amplifier crystal plates, is h=2×10$^{-3}$ m. The bulk velocity used for this study is $U_b$=10 m/s, although typical velocities in laser amplifiers could be 5 times higher. The flow is assumed to be turbulent and fully developed. The corresponding bulk Reynolds number $Re_b = \frac{U_b\, 2h}{\nu} = 14000$ corresponds to a friction Reynolds number of $Re_\tau = \frac{u_\tau h}{\nu} = 400$.

Treating turbulence by LES consists in resolving the large scale of turbulence while modelling the small scales. The scales are separated by filtering the conservation equations. A detailed description of this type of simulations can be found in [12]. The WALE subgrid scale model, developed by Nicoud & Ducros [13] is used as it is well adapted for wall-bounded flows. Wall functions are not used.

The energy of the laser is accumulated in the amplification crystals and evacuated by the helium flow. The heat balance of the flow shows that the mean temperature increase of the fluid between the inlet and the outlet of the domain is below 1.3 K. The impact of temperature variations on physical properties is neglected as it is of 1 to 2 %. The fluid properties are taken as constant and the temperature is considered as a passive scalar (Pr=0.70). The used fluid and solid properties are summarized in Table 1.

**Table 1** Constant physical properties of fluid and solid

| | Fluid | Solid |
|---|---|---|
| Kinematic viscosity ν (m$^2$/s) | 3×10$^{-6}$ | - |
| Thermal conductivity λ (W/mK) | 0.064 | 35 |
| Heat capacity $C_p$ (J/kgK) | 5205 | 100 |
| Density ρ (kg/m$^3$) | 2.982 | 4583 |

For the fluid, the governing set of conservation equations in the filtered form, represented by $\widetilde{()}$, are then for an incompressible flow: the conservation of mass (1), momentum (2) and energy (3). For the solid, the conservation of the energy is given in equation (4):

$$\frac{\partial \widetilde{u_i}}{\partial x_i} = 0 \quad (1)$$

$$\frac{\partial \widetilde{u_i}}{\partial t} + \frac{\partial \widetilde{u_i}\widetilde{u_j}}{\partial x_j} = \frac{-1}{\rho_f}\frac{\partial \widetilde{P}}{\partial x_i} + \nu \frac{\partial}{\partial x_j}\left(\frac{\partial \widetilde{u_i}}{\partial x_j} + \frac{\partial \widetilde{u_j}}{\partial x_i}\right) \quad (2)$$

$$\frac{\partial \widetilde{T}}{\partial t} + \frac{\partial \widetilde{u_i}\widetilde{T}}{\partial x_i} = \alpha_f \frac{\partial^2 \widetilde{T}}{\partial x_i^2} + q_f \quad (3)$$

$$\frac{\partial T}{\partial t} = \alpha_s \frac{\partial^2 T}{\partial x_i^2} + q_s \quad (4)$$

The heat deposited by the pumping is taken into account via the term $q = \frac{Q}{C_p \times \rho \times V}$ where Q is the heat source.

A 3$^{rd}$ order Runge-Kutta scheme is used for time integration. The discretization in space is done by the MAC method of Harlow & Welsh [14], using 2$^{nd}$ order centered diffusion schemes for momentum and energy. For dynamic convection, a 2nd order centered scheme is used. For thermal convection, a 3rd order (QUICK) scheme is applied.

## RESULTS OF THE NUMERICAL ANALYSIS

The results of the LES calculations are presented as non-dimensional quantities, normalized using the classical scaling method [12]:

$$U^+ = \frac{\langle U \rangle}{U_\tau}, \quad U_{i\,RMS}{}^+ = \frac{U_{i\,RMS}}{U_\tau}, \quad y^+ = \frac{y\,U_\tau}{\nu}, \quad T^+ = \frac{T_w - \langle T \rangle}{T_\tau}, \quad T_{rms}{}^+ = \frac{T_{rms}}{T_\tau} \quad (5)$$

The velocities are normalized by the friction velocity, calculated from the wall normal gradient of the mean velocity:

$$U_\tau = \sqrt{\nu \left(\frac{\partial \langle U \rangle}{\partial y}\right)_w} \quad (6)$$

The temperatures are normalized by the friction temperature, calculated from the wall normal gradient of the mean temperature and the friction velocity:

$$T_\tau = \frac{\lambda \left(\frac{\partial \langle T \rangle}{\partial y}\right)_w}{\rho C_p U_\tau} \quad (7)$$

**Heated periodic channel flow**

In the first step, the method to simulate conjugated heat transfer in a laser amplifier is validated by simulating by LES a streamwise (x) and spanwise (z) bi-periodic plane channel that is bounded crosswise (y) by two heated solid plates. The computational domain is shown in Fig. 2. The dimensions of the channel are $L_x \times L^f_y \times L_z = 4\pi h \times 2h \times 2\pi h$. Both plates have the streamwise and spanwise dimensions of the channel; their heights are $L^s_y = 3.5 \times 10^{-3}$ m. The fluid mesh contains $N_x \times N^f_y \times N_z = 120 \times 192 \times 240$ cells. The cells are streamwise and crosswise homogeneously distributed with non-dimensional sizes of $\Delta_x^+ = 40$ and $\Delta_z^+ = 10$, scaled by the friction velocity $u_\tau$. The mesh is refined crosswise near the wall to $\Delta_y^+ = 1$. Starting at the wall, the cells are gradually coarsen crosswise following a tangential hyperbolic law. The maximum cell size at the center of the channel is $\Delta_y^+ = 15$. The solid mesh contains $N^s_y = 30$ homogeneously distributed cells.

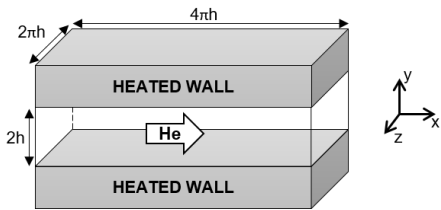

**Figure 2** Computational domain with flow channel and heated plates

Statistics are collected by averaging over the two homogeneity directions x and z, and over time. 0.1 s are simulated for the statistics, which represents 30 characteristic times defined as $h/U_\tau$.

Fig. 3 presents the crosswise profile of the normalized streamwise mean velocity $U^+$ plotted with the normalized distance to the wall $y^+$. The comparison shows a satisfying agreement with DNS results of Kawamura [7] realized at the same Reynolds number although the velocity is under-estimated in the center of the channel. This result can be improved by using a 4th order scheme for dynamic convection; the 2nd order scheme was preferred for stability reasons.

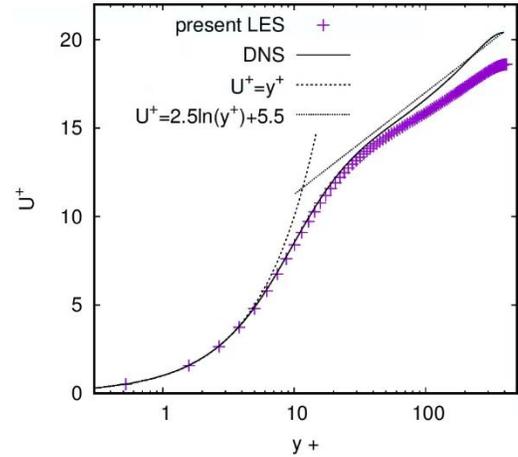

**Figure 3** Normalized mean streamwise velocity in the channel flow, compared to DNS data of Kawamura *et al.* [7]

Fig. 4 presents in the same way the normalized velocity fluctuations $U_{iRMS}^+$ in the three coordinate directions, which are compared to the same DNS of Kawamura [7]. This comparison also shows a good agreement between DNS and LES.

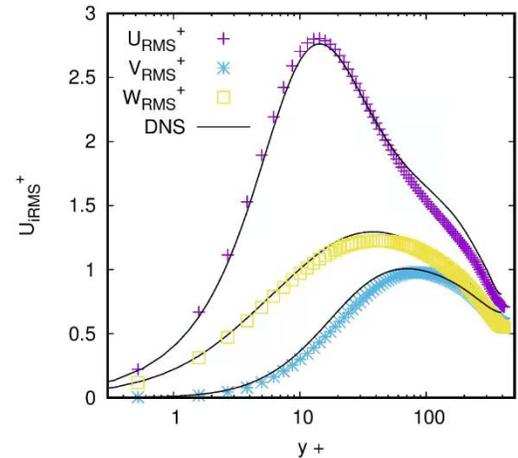

**Figure 4** Normalized root mean squares of velocity in the channel flow, compared to DNS data of Kawamura *et al.* [7]

In a first approach, a total thermal power of $Q_s = 9.2$ W is deposited in the solid plates by a constant volume heat source. $Q_s$ is representative of a cryogenically cooled multi-slab amplifier [15], proportionally to the simulated volume. In order to conserve the energy in the calculation domain, the same thermal power is removed from the fluid by a constant volume source term ($Q_f$= -$Q_s$). The initial temperature of the solid has a constant value of $T_s$=105 K and of the fluid of $T_f$=80 K.

In an actual laser, the thermal power is not deposited constantly in the amplifier but only during discrete nanosecond pulses with a typical frequency of 10 Hz [15]. The thermal balance in the solid is as follows:

$$m_s C_{ps} \frac{dT_s}{dt} = Q_s - S \times h_{exch} \times (T_s - T_f) \quad (8)$$

Where $m_s$= ($\rho \times L_x \times L_y \times L_z$)$_s$ is the solid mass, S=$L_x \times L_z$ is the exchange surface and $h_{exch}$ is the heat transfer coefficient. This leads to a thermal time constant:

$$\mathcal{T} = \frac{(\rho \times C_p \times L_y)_s}{h_{exch}} \quad (9)$$

The heat transfer coefficient can be approximated to $h_{exch}$=600 W/m$^2$K using the classical Dittus-Boelter correlation: Nu=0.023Re$^{0.8}\times$Pr$^{0.4}$ [16]. The thermal time constant is then estimated to $\mathcal{T}$=2.6 s. In other words, in 0.1 s (the time between two laser pulses) the solid temperature decreases by about 1 K, less than 4% of the fluid/solid temperature difference. These variations are here neglected by the constant heat source approximation and the simulation reaches a thermal equilibrium. 0.2 s were simulated to attain the thermal equilibrium prior to the 0.1 s simulated to collect the statistics.

Fig. 5 and 6 respectively present the normalized mean temperature $T^+$ and normalized the root mean square of temperatures $T_{RMS}^+$ plotted with the normalized distance to the wall $y^+$. They are again compared to DNS results of Kawamura [7]. The mean temperature profile shows an excellent agreement between LES and DNS while the temperature fluctuations are underestimated.

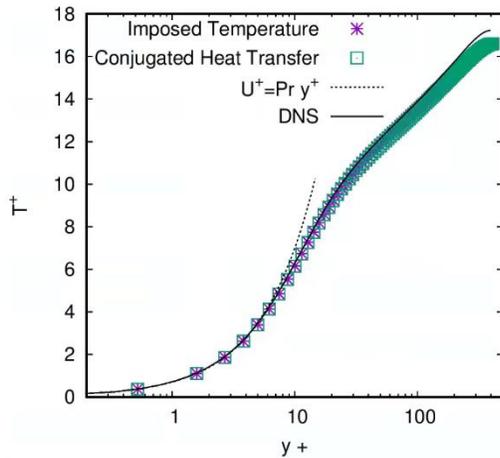

**Figure 5** Normalized mean temperature in the heated periodic channel flow with and without plates, compared to DNS data of Kawamura *et al.* [7]

The temperature fluctuations in the fluid tend to zero at the wall, while temperatures barely fluctuate in the solid plates: the temperature root mean squares range from 2×10$^{-2}$ K at the fluid interface to 3×10$^{-4}$ K at the opposite face. This is coherent with the fluid and solid characteristics: the thermal activity ratio $K = \sqrt{(\lambda \rho C_p)_f / (\lambda \rho C_p)_s} = 8.10^{-3}$ is small, suggesting that the fluid/solid interface behaving similarly to an ideal non-fluctuating temperature boundary condition [9].

To confirm the analogy, the same fluid channel was simulated using a Dirichlet boundary condition with imposed temperature of $T_w$=104.5 K at the wall. $T_w$ corresponds to the wall temperature of the solid plates in the conjugated heat transfer case. The results of the imposed temperature simulation are also plotted in Fig. 5 and 6. Both mean and fluctuating temperatures perfectly collapse with the results of the conjugated heat transfer simulation.

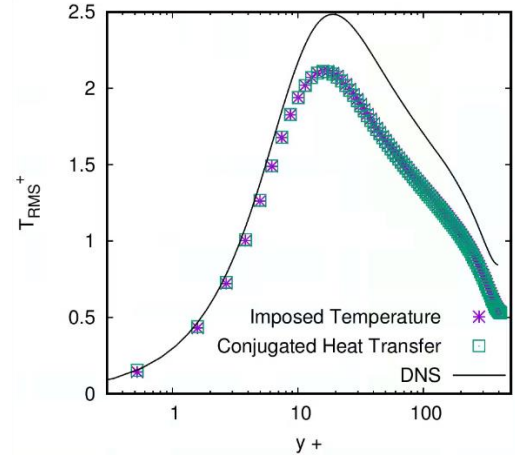

**Figure 6** Normalized root mean square of temperature in the heated periodic channel flow with and without plates, compared to DNS data of Kawamura *et al.* [7]

**Heated open channel flow**

In the actual amplifier, boundary layers will develop along the channel, which may affect the laser beam quality. Indeed, different parts of the laser beam will propagate through a different thickness of thermal boundary layer, and will therefore meet different refractive index variations, affecting their optical paths differently. If the optical paths difference is of the order of or higher to the wavelength, the laser beam loses its coherence.

The results obtained in a streamwise periodic channel flow do not give information on this thermal boundary layer development, and it is necessary to study a thermally open channel. The dynamic of the flow is however kept as a fully developed turbulent plane channel. To do so, a periodic isothermal channel is used as precursor to supply the entrance instantaneous velocity field, as shown in Fig. 7. The mesh characteristics are the same as in the previous case of a periodic channel simulation.

The previous validation simulations presented no difference in the temperature fluctuations between the conjugated heat transfer and the imposed temperature boundary condition simulations, for fluid and solid characteristics set to typical values of a laser amplifier. The open channel flow simulation was therefore conducted using an imposed temperature boundary condition at the walls.

The wall temperature distribution was obtained by a RANS simulation of an open channel coupled with two amplifier plates. The wall temperature is spanwise homogeneous, and it varies streamwise as follows:

$$T_w = 0.017 \times (x/h)^2 + 0.46 \times x/h + 100 \quad (10)$$

The inlet fluid temperature is homogeneous at 80 K.

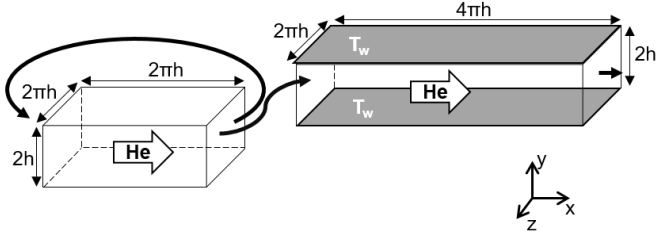

**Figure 7** Computational domain with precursor channel and heated open channel

Since the open channel is not streamwise homogeneous, the statistics are now collected by averaging spanwise and over time. Starting from the RANS solution, 0.05 s are simulated prior to 0.2 s (*i.e.* 60 characteristic times) for the statistics computation.

Fig. 8 and 9 respectively present the mean temperature T and the root mean square of temperatures $T_{RMS}$ on a (xy) surface, thus highlighting the thermal boundary layers developments along the heated open channel. The helium flow enters the channel at T=80 K from the left and it is heated by the upper and lower walls as it flows to the right.

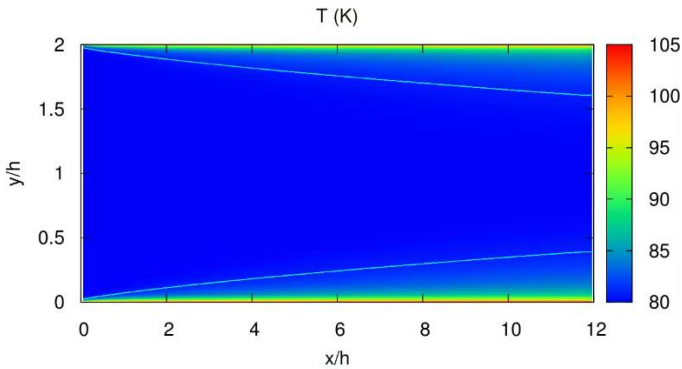

**Figure 8** Mean temperature on a streamwise plane in the heated open channel flow, with contour line at constant temperature T=81 K

The root mean square of temperature reach a maximum 3.5 K on streamwise strips close to the walls, where are also localized the maximum of velocity fluctuations. The temperature fluctuations widen toward the center of the channel as the thermal boundary layers thicken. It should be reminded from the validation simulations that the temperature fluctuations peaks may be underestimated.

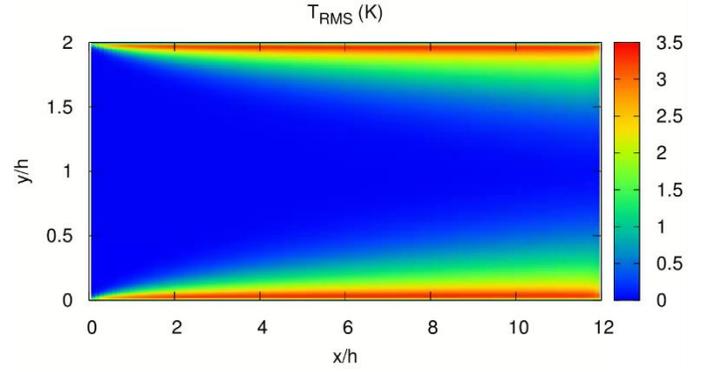

**Figure 9** Root mean square of temperature on a streamwise plane in the heated open channel flow

According to the Gladstone-Dale law, the index of refraction of a gas varies with its density following:

$$n(x, y, z) = 1 + G \times \rho(x, y, z) \quad (11)$$

Where G is a constant. The helium temperature variations induce density variations and thus index of refraction variations shown Fig. 10. The index of refraction variations are small, confirming the choice of helium as coolant. Such small variations will not affect the laser beam amplitude, but they may induce a phase distortion.

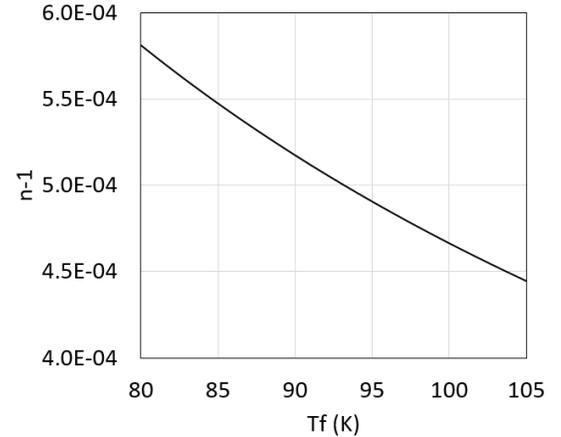

**Figure 10** Index of refraction variations in gaseous helium at P=5.10$^5$ Pa and at low temperatures

The phase variation of a beam propagating crosswise through the channel flow is:

$$\Delta\phi(x,z) = k \int_0^{2h} \Delta n(x,y,z) dy \quad (12)$$

Where $\Delta n(x, y, z) = n(x, y, z) - n(0, y, z)$ is the variation of the refractive index compared to one at the inlet of the channel; and k=2π/λ$_l$ is the wave number. For a Yb:YAG laser, the wavelength is λ$_l$=940×10$^{-9}$ m.

The mean temperature field is spanwise homogeneous in the simulated open heated channel. The phase variation induced by the mean temperatures varies monotonously between the inlet and the outlet of the channel. The maximum phase variation is then $\Delta\phi(12h) = 0.3$ rad.

It should be noted that this phase variation of 0.3 rad takes into account only one channel of the multi-slab configuration, and that only about a quarter of the actual channel length is here simulated. Moreover, only the mean temperature variations were considered, and the temperature fluctuations influence remains to be investigated.

## CONCLUSION

Large Eddy Simulations were conducted in heated turbulent channel flows of low temperature gaseous helium. In a first step, the channel was streamwise periodic and it was crosswise bounded by two solid plates, using conjugated heat transfer. It was shown that for the physical characteristics representative of a cryogenically cooled laser amplifier, the mean and fluctuating temperatures results were similar to an ideal imposed temperature boundary condition simulation. To observe the thermal boundary layers developments, an open heated channel flow was then simulated using such wall thermal boundary condition. A first evaluation of the thermal boundary layer impact on the laser beam coherence was conducted for mean values. Large Eddy Simulation also gives access to the temperature fluctuations. They were computed along the heated channel, but their effect on the laser beam remains to be investigated. The next step will then consist in taking into account the dynamic boundary layers developments.

## AKNOWLEDGEMENTS

This work was granted access to the HPC resources of CINES under the allocation 2019-A0052A10626 attributed by GENCI (Grand Équipement National de Calcul Intensif).

We acknowledge the financial support of the Cross-Disciplinary Program on Numerical Simulation of CEA, the French Alternative Energies and Atomic Energy Commission.

## REFERENCES

[1] B. Le Garrec, "Laser-diode and Flash Lamp Pumped Solid-State Lasers," in *AIP Conference Proceedings*, 2010, vol. 1228, pp. 111–116.

[2] V. Cardinali, E. Marmois, B. Le Garrec, and G. Bourdet, "Thermo-optical measurements of ytterbium doped sesquioxides ceramics," in *Proceedings of SPIE - The International Society for Optical Engineering*, 2010, vol. 7721, p. 77210U.

[3] D. Brown *et al.*, "The Application of Cryogenic Laser Physics to the Development of High Average Power Ultra-Short Pulse Lasers," *Appl. Sci.*, vol. 6, no. 1, p. 23, Jan. 2016.

[4] J. P. Perin, F. Millet, B. Rus, and M. Divoký, "Cryogenic Cooling For High Power Laser Amplifiers," presented at the 5th International Conference On The Frontiers Of Plasma Physics And Technology, Singapore, 2011.

[5] E. Tromeur, E. Garnier, and P. Sagaut, "Large-eddy simulation of aero-optical effects in a spatially developing turbulent boundary layer," *J. Turbul.*, vol. 7, p. N1, Jan. 2006.

[6] A. Mani, M. Wang, and P. Moin, "Resolution requirements for aero-optical simulations," *J. Comput. Phys.*, vol. 227, no. 21, pp. 9008–9020, Nov. 2008.

[7] H. Kawamura, H. Abe, and Y. Matsuo, "DNS of turbulent heat transfer in channel flow with respect to Reynolds and Prandtl number effects," *International Journal of Heat and Fluid Flow*, pp. 196–207, 1999.

[8] L. Bricteux, M. Duponcheel, G. Winckelmans, I. Tiselj, and Y. Bartosiewicz, "Direct and large eddy simulation of turbulent heat transfer at very low Prandtl number: Application to lead–bismuth flows," *Nucl. Eng. Des.*, vol. 246, pp. 91–97, May 2012.

[9] I. Tiselj and L. Cizelj, "DNS of turbulent channel flow with conjugate heat transfer at Prandtl number 0.01," *Nucl. Eng. Des.*, vol. 253, pp. 153–160, Dec. 2012.

[10] E. Merzari, W. D. Pointer, J. G. Smith, A. Tentner, and P. Fischer, "Numerical simulation of the flow in wire-wrapped pin bundles: Effect of pin-wire contact modeling," *Nucl. Eng. Des.*, vol. 253, pp. 374–386, Dec. 2012.

[11] P.-E. Angeli, U. Bieder, and G. Fauchet, "Overview of the TrioCFD code: main features, V&V procedures and typical applications to nuclear engineering," in *Proceedings of 16th International Topical Meeting on Nuclear Reactor Thermal Hydraulics (NURETH-16)*, 2015.

[12] P. Sagaut, *Large Eddy Simulation for Incompressible Flows*. Berlin, Heidelberg: Springer Berlin Heidelberg, 2002.

[13] F. Nicoud and F. Ducros, "Subgrid-scale stress modelling based on the square of the velocity gradient tensor," *Flow Turbul. Combust.*, vol. 62, no. 3, pp. 183–200, 1999.

[14] F. H. Harlow and J. E. Welch, "Numerical Calculation of Time-Dependent Viscous Incompressible Flow of Fluid with Free Surface," *Phys. Fluids*, vol. 8, no. 12, p. 2182, 1965.

[15] M. Divoky *et al.*, "Conceptual design of 100 J cryogenically-cooled multi-slab laser for fusion research," *EPJ Web Conf.*, vol. 59, p. 08004, 2013.

[16] F. W. Dittus and L. M. K. Boelter, "Heat transfer in automobile radiators of the tubular type," *Int. Commun. Heat Mass Transf.*, vol. 12, no. 1, pp. 3–22, 1985.